# Superconducting Vortices induced Periodic Magnetoresistance Oscillations in Single Crystal Au Nanowires

Lin He,[*,1,2] Jian Wang,[*,2,3]

[1] Department of Physics, Beijing Normal University, Beijing, 100875, People's Republic of China

[2] The Center for Nanoscale Science and Department of Physics, The Pennsylvania State University, University Park, Pennsylvania 16802-6300, USA

[3] International Center for Quantum Materials and State Key Laboratory for Mesoscopic Physics, School of Physics, Peking University, Beijing, 100871, People's Republic of China.

*Corresponding author email address: helin@bnu.edu.cn and jianwangphysics@pku.edu.cn

## ABSTRACT

We show in this paper that it is possible to induce superconducting vortices in a gold nanowire connected to superconducting electrodes. The gold nanowire acquires superconductivity by the proximity effect. The differential magnetoresistance of the nanowire beyond a critical magnetic field shows uniform oscillations with increasing field with a period of $\Phi_0/(2\pi r^2)$ ($\Phi_0 = h/2e$ is the superconducting flux quantum, $r = 35$ nm is the radius of the nanowire). We demonstrate that these periodic oscillations are the signatures of the sequential generation and moving of vortices across the gold nanowire.

Vortices are excluded from a superconductor at low magnetic field. When the field is increased beyond the critical value, vortices proliferate and superconductivity is destroyed.[1-3] A great deal of effort has been spent to image and to manipulate an individual vortex[4-8] and to image array of vortices.[1-3] In addition to the fascinating physics, these efforts are motivated by the possible incorporation of individual vortex in a fluxon-based electronic circuit.[9-12] An interesting and fundamental question about vortices to inquire is to find out if it is possible to induce superconducting vortices in a normal nanowire by proximity to superconductors. Recently, Cuevas and Bergeret studied this question theoretically and acquired a positive answer.[13] In this paper we show evidence of the generation of superconducting vortices in a gold nanowire connected to superconducting electrodes. In quasi-one-dimensional superconductors, the superconducting vortices generated by perpendicular magnetic field are usually along the wire.[14-17] In our superconductor-Au nanowire-superconductor (S-NW-S) structure, the magnetoresistance of the Au wire shows up phase slips, which arise from the generation and moving of vortices across the wire. The proximity induced superconductivity in the Au wire allows for generation of individual vortices one at a time along the nanowire by gradually increasing the perpendicular magnetic field uniformly beyond a critical value. As a result, the resistance of the Au wire increases with increasing the field step by step for that the generated vortices continuously move across the wire.

Recently, it is shown that single crystal Au nanowires can acquire superconductivity via the proximity effect of superconducting electrodes.[18,19] In a previous paper, we reported on experimental studies of proximity effect in S-NW-S structures.[18] The experimental configuration is shown in Fig. 1(a). Four focused ion beam (FIB)-assisted superconducting W electrodes were deposited to contact individual Au nanowires of 70 nm diameter for standard four-probe transport measurements. Fig. 1(b) shows the high-resolution transmission electron microscope image and the corresponding selected-area electron diffraction (SAED) patterns of the Au nanowire. These images indicate that the Au nanowire is of high quality single crystal. The W strips used as the electrodes are composed of tungsten, carbon and gallium and show a superconducting transition temperature of ~ 5.1 K.[18,20,21] The Au nanowires acquire superconductivity via the proximity effect from the W electrodes. The superconducting gap function decreases gradually as one moves from the S-NW interface towards the middle of the nanowire.[18,22] A wire of 1.0 μm in length was found to be completely superconducting below 4.1 K. A 1.2 μm wire showed superconducting drop in two steps near 4.1 and 3.5 K achieving zero resistance below 3.4 K. A minigap phase, which is a state separating the normal and superconducting regions,[23,24] of the 1.2 μm wire was found between 2.5 and 3.4 K.[18] Below 4.1 K, the resistance of the nanowire $R$ increases with the applied field $B$. However the increase is not uniform, instead, $R$ vs $B$ shows terraces for measurements made between 2 and 3.5 K. There is no hysteresis in the terraces of $R$ vs $B$ curves and the $R$ vs $B$ curves are symmetrical about zero applied field. Upon differentiation, the terraces in $R$ vs $B$ show up as periodic oscillations in ($dR/d|B|$) with a uniform period in magnetic field of 0.25 T.[18] The $dR/d|B|$ results at different temperatures of both the 1.0 and 1.2 μm wires are shown in Fig. 2. The results at 3 K for positive magnetic field are magnifield and reproduced in Fig. 1(c). Except for the first peak of the 1.2 μm nanowire,[18] 15 peaks in $dR/d|B|$, starting from about 1.65 and 1.6 T and separated by 0.25 T, can be resolved respectively for the 1.0 and 1.2 μm wires. The nature of the periodic oscillation was not fully understood and not explained in reference 18. We present evidence here to show that these oscillations are very likely signatures of sequential generation and moving of superconducting vortices in response to an increasing perpendicular magnetic field threading the nanowire. The staircase like resistance is also accounted well in our model.

The magnetic field value for the first (second) and subsequent peaks in $dR/d|B|$ for the 1.0 μm (1.2 μm) wire are temperature dependent. The magnetic fields for every peak of the 1.0 μm wire at 2, 2.5, 3 and 3.5 K are shown in Fig. 3(a). The linear relation between the index number of the peaks ($N$) and the temperature dependent magnetic field value ($B$) of the 1 μm wire, as shown in Fig. 2(a), can be expressed as

$$\Delta B\ (N-1) = (B - B_p(T)). \qquad (1)$$

Here, $B_p$ is the temperature dependent magnetic field of the first peak in the $dR/d|B|$ curves and $\Delta B$ is the magnetic field step between neighboring peaks (A similar expression can be written for the 1.2 μm wire for peaks with $N \geq 2$. The physics to be presented for the 1 μm wire can also be extended to the 1.2 μm wire. For clarity we will limit our discussions from this point onto the 1 μm wire.). The first peak of the 1.2 μm wire appears at a temperature independent field value of $B \sim 0.25$ T. The physical mechanism responsible for the appearance is related to the minigap phase and different from the onset of the other peaks in the 1.2 μm and the 1 μm wire. At $B \sim 0.25$ T, the enclosed magnetic flux in the 1.2 μm nanowire is close to $\Phi_0/2$. If one vortex is generated in the wire, then the vortex is surrounded by clockwise or anticlockwise superconducting current to reducing or enhancing the flux to 0 or $\Phi_0$ respectively.[10,11] The interesting phenomena at about $\Phi_0/2$ of the 1.2 μm wire may arise from the quantum superposition of flux quanta 0 and $\Phi_0$ trapped in the system. Experiments and analysis to explore the nature of this peak are carrying out and will be the subject of a separate publication. Fig. 3(b) shows (more clearly than Fig. 3(a)) that $B_p$ decreases linearly with increasing temperature and vanishes at about 3.9 K. This is consistent with the experimental result that no oscillation in $dR/d|B|$ is seen at and above 4 K.[18] On the other hand, the slope $\Delta B = (B-B_p)/(N-1)$ appears to be temperature independent with a value of $0.25 \pm 0.015$ T. We note that if we divide the superconducting flux quantum, $\Phi_0 = h/2e$, by the area of $2\times(\pi r^2)$, with $r = 35$ nm being the radius of the gold wire, the corresponding magnetic field is about 0.25 T.

We now turn to understand the linear behavior of the number of peaks $N$ vs the applied field $B$ in the S-NW-S structures. In a careful experimental imaging vortices near the critical temperature in superconducting Nb strips with widths of 1.6 μm, 10 μm and 100 μm, it is found experimentally that the dependence of the number of vortices on the magnetic field at high field region can be expressed approximated by $N = (B-B_m)A/\Phi_0$, where $A$ equals to the total area of the strips and $B_m$ correspond to a critical field for generating the first vortex in the strips.[15] This indicates that the number of vortices $N$ increases linearly with the magnetic field. This relation is identical to equation (1) suggesting that the peaks of the differential magnetoresistance in our gold wire likely arise from the generation of superconducting vortices. Due to the large area of the strips in reference 15, the critical field is small and the number of vortices trapped in the strips increase very quickly with the applied field.

For our S-NW-S structure, the proximity induced superconducting gap in the normal metal nanowire in zero magnetic field can be qualitatively expressed as $\Delta(x) = \Delta_s \cosh(x/\xi_N)/\cosh(L/2\xi_N)$, where $L$ is the length between two superconducting electrodes of the nanowire ($-L/2 < x < L/2$), $\xi_N$ is the coherence length characterizing the decay of the superconductivity in the nanowire, $\Delta_s$ is the gap at the S-NW boundary.[18,22] The superconducting gap decreases gradually from $\Delta_s$ at the S-NW boundary to a smaller one $\Delta(x = 0) = \Delta_s/\cosh(L/2\xi_N)$ in the middle of the nanowire, as shown in Fig. 4(a). The gap function $\Delta(x = 0)$ as well the critical field $B_C$ required to drive the central part of the wire normal are temperature dependent. By gradually increasing the magnetic field perpendicular to the nanowire, the central part of the nanowire will first be driven normal allowing one vortex carrying one flux quantum $\Phi_0$ enters the nanowire, as shown in Fig. 4(b). The field needed to generate first vortex ($N = 1$), according to equation (1) is $B_p$. It is reasonable to express $B_p$ as a sum of $B_C + B_1$, where $B_1$ is the magnetic field for generating the first vortex when the wire is at the edge of being normal. Since the cross-sectional area for the first vortex threading the wire is $\sim \pi r^2$. $B_1 = \Phi_0/(\pi r^2)$ is estimated as a numerical value of 0.5 T. When the field is further increased to $B_2 = B_p + \Delta B_V$ to allow the second vortex to condense into the nanowire, the additional field, $\Delta B_V$ is equal to $\Phi_0/(2\pi r^2)$ or 0.25 T.[25] Then the field for the third vortex in the wire is $B_3 = B_p + 2\Delta B_V$ and the trapped flux is $(B_1+2\Delta B_V)\times(3\pi r^2) = 6\Phi_0$. In general, we obtain

$$B_N = B_p + (N-1)\Delta B_V, \qquad (2)$$

where $B_N$ is the magnetic field for generating $N$ vortices. The total trapped flux in the wire with $N$ vortices is $[N(N+1)/2]\Phi_0$. We can see that equation (2) is identical to equation (1), which supports the interpretation that the peaks in $dR/d|B|$ are signatures of the sequential appearance of vortices in the Au nanowire. In bulk superconductors, one vortex usually can only carry a single flux quantum. In small superconductors, however, both experiments[26] and theory[27] pointed out that vortex can hold multiple flux quanta, which are called as giant vortex state. Our experiment and analysis indicate that the vortex generated in the S-NW-S structure is similar as the giant vortex observed in mesoscopic superconductors.

Below the critical field $B_p$, all flux was expelled from the nanowire, the resistance of the nanowire is zero; above this field vortices are induced one by one with magnetic field widths of $\Phi_0/(2\pi r^2)$. The vortices on the Au wire can act as phase slip centers. Then the resistance can arise from the phase slip process induced by thermal fluctuation or quantum fluctuation, which is usually observed in the destruction of superconductivity in quasi-one-dimensional systems.[28-31] In the S-NW-S structure, we demonstrate below that the dissipation arises mainly from the generation and motion of vortices across the Au wire rapidly and continuously due to the Lorentz force acting on them.[31] A transport supercurrent $I_s$ along the wire exerts a Lorentz force $F_L = I_s \times [N(N+1)/2]\Phi_0$ on the vortices and the motion is opposed by a viscous drag $F_d = -\eta v$, where $\eta$ is the viscosity coefficient of vortices and $v$ is the velocity of the vortices.[32] The motion of $N$ vortices across the wire generates a voltage drop $V_N = L \times [N(N+1)/2]\Phi_0 \times v$ along the wire. Therefore, the supercurrent transfers through the Au wire is $I_s = (I - V_N/R_n)$, where $I$ is the total applied current and $R_n$ is the resistance of the wire in normal state. Then we can express the above analysis as

$$(I - V_N/R_n)[N(N+1)/2]\Phi_0 = \eta v. \qquad (3)$$

By solving equation (3) and $V_N = L \times [N(N+1)/2]\Phi_0 \times v$ together, we obtain the voltage drop arising from the motion of $N$ vortices as

$$V_N = [N(N+1)]^2 I / [a + \frac{[N(N+1)]^2}{R_n}], \qquad (4)$$

or the resistance due to the move of the $N$ vortices across the wire as $R_N = V_N/I$. In equation (4), $a = 4\eta/(L\Phi_0^2)$. In our S-NW-S structure, $R_n = 197\ \Omega$ and the resistance of $R_{15}$ is about 190 $\Omega$. According to equation (4), we obtain $a = 14\ \Omega^{-1}$ in the Au wire. Figure 5(a) shows the calculated resistance $R_N = V_N/I$ by taking into account $a = 14\ \Omega^{-1}$ and $R_n = 197\ \Omega$. The experimental result measured at 3 K was also shown for comparison. Figure 5(b) shows the height of the peaks in the $dR/d|B|$ curve as a function of the number of vortices measured at $T = 3$ K for positive field. The height of $dR/d|B|$ curves, as shown in Fig. 5, reflects the increase of resistance for the generation and motion of the $N^{th}$ vortex $\Delta R_N$ because the field width from ($N$-1) vortices to $N$ vortices is a constant $\Delta B_V$. The $dR/d|N|$ calculated by the theoretical curve in Fig. 5(a) is also shown in Fig. 5(b). Obviously, our model accounts well for the experimental data.

The number of the peaks of the $dR/d|B|$ curves, about 15 peaks for the 1.0 μm wire (as shown in Fig. 1(c)), indicates that it is determined by the largest number of vortices ~ $L/2r$ in the nanowire. It is interesting that if we assume the spatial extent of a vortex along the length of the wire is also $2r$, then a 1.0 μm wire can indeed accommodate a maximum of about 15 vortices when it is under a magnetic field of $B_{15}$ of ~ 5 T. Under such a condition, the 1 μm Au nanowire is in the normal state, possibly as a consequence of the 'melting' of the vortex structure in the wire.[3]

Figure 6 (a) and (b) show the dependence of voltage drop on the number of vortices in the S-NW-S junction calculated by Eq. (4). In the calculation, we assume $L/2r$ ~ 15 ($L$ is the length of the nanowire between V+ and V- electrodes), i.e., the largest number of vortices ~ $L/2r$ generated in the nanowire is 15. By increasing the viscosity coefficient of vortices $\eta$ (or the resistance of the nanowire in the normal state $R_n$), the voltage drop generated by the motion of vortices across the wire increases more and more slowly with the number of vortices. According to the model, the dependence of voltage drop on the positive and negative magnetic field should be symmetry and no hysteresis should be observed. These results are in good agreement with our experimental results. Upon differentiation of $V_N$ calculated by Eq. (4), we can obtain $dV/dN$ curves as a function of the number of vortices with various $R_n$ and $a$. The height of $dV/dN$ curves, as shown in Fig. 5 for a typical case (The $V_N$ are calculated with $R_n$ = 197 Ω and $a$ = 14 Ω$^{-1}$), reflects the increase of voltage for the generation and motion of the $N^{th}$ vortices. Obviously, the increase of voltage depends on the number of vortices and there is a peak in the $dV/dN$ curve. We calculated $dV/dN$ curves as a function of the number of vortices with various $R_n$ and $a$. It is interesting to note that the position of the peak in the $dV/dN$ curves depends logarithmic on the parameters $R_n$ and $a$ of the S-NW-S junction, as shown in Fig. 6(c). It indicates that the parameters $R_n$ and $a$ have similar effects on the voltage staircase pattern of the S-NW-S junction, or the parameters $R_n$ and $a$ play similar role in determining the transport properties of the S-NW-S junction.

Figure 7 shows the calculated vortices velocity as a function of the number of vortices with various $R_n$ and $a$. For small viscosity coefficient or small normal state resistance, the vortices velocity increases with increasing the number of vortices until reaching a maximum. Then the velocity decreases with further increasing the number of vortices. For sufficient large viscosity coefficient or normal state resistance, i.e., $a = 100$ $\Omega^{-1}$ or $R_n = 2000$ $\Omega$, the vortices velocity increases monotonic with increasing the number of vortices to 15. For the S-NW-S junction reported in this work, the total applied current $I = 100$ nA, $a = 14$ $\Omega^{-1}$, and $R_n = 197$ $\Omega$, the velocity of the first vortex is estimated as 0.0285 m/s. It increases to the maximum ~ 0.374 m/s when the generated number of vortices is 7. Then velocity decreases to 0.156 m/s with the number of vortices reaching 15. According to our analysis, the voltage drop (or resistance $V_N/I$) staircase pattern of the S-NW-S junction is determined by two key factors, i.e., the viscosity coefficient of vortices $\eta$ and the resistance of the nanowire in the normal state $R_n$. The parameters $R_n$ and $\eta$ play similar role in determining the transport properties of the S-NW-S junction.

For $N$ vortices induced in the nanowire by the magnetic field, the trapped flux quanta are $[N(N+1)/2]\Phi_0$, as shown in Fig. 4(b). Then the average flux quanta per vortex are $[(N+1)/2]\Phi_0$. When $N$ is a odd number, each vortex carries equal integral flux quanta. When $N$ is even, the average flux quanta per vortex are not integer. Let's discuss the simplest case, *i.e.*, $N = 2$ and the total trapped flux quanta are $3\Phi_0$. Due to the symmetry of the S-NW-S structure under consideration, the magnetic flux trapped in the wire should be uniform. Then the two vortices should have equal probability to trap one flux quantum or two flux quanta. This suggests that there is a sequence of frequent discrete transitions of the trapped flux quanta between $2\Phi_0$ and $\Phi_0$ in the vortices due to the flux quantization. According to Heisenberg's uncertainty relations, the vortex's charge and the magnetic flux form a conjugate pair, with minimum uncertainty $\Delta Q \Delta \Phi \sim h$. With considering $\Delta \Phi \sim \Phi_0$, we have $\Delta Q \sim 2e$. It appears there is the motion of a Cooper pair with charge $2e$, from one vortex to the other that is accompanying with the transport of a flux quantum among vortices. Recently, phase slips induced by Coulomb interactions of electrons are measured in a chain of Josephson junctions.[33,34] The chain of vortex in the nanowire of the S-NW-S structure, which is similar as a Josephson junction chain, may be served as model system for accessing the quantum-level phase slip process in superconductors.

There are two vital prerequisites to observe the generation and annihilation of superconducting vortex one by one in the S-NW-S structure. First, our experimental set-up (the S-NW-S structure) provides a system where multiple vortices can be accommodated one by one along the nanowire. Second, the critical field for the superconductor-normal metal transition of the FIB-deposited W electrodes in our S-NW-S structures is large. For example, the onset critical field of the 1 μm nanowire is 6.2 T at 2 K. Furthermore, the diameter of the nanowire is also a good match to observe a large number of resistance staircases in our system.

Finally, we note that there is a crucial difference between the vortices pattern observed in our S-NW-S structure and that predicted in theory.[13] In our S-NW-S structure, the vortices appear to form a one-dimensional chain along the wire. However, in reference 13, the vortices were predicted to form one-dimensional structure perpendicular to the junction. Cuevas and Bergeret calculated vortices structures in the normal metal with $W \geq L = \xi$, here $W$ and $L$ are the width and length of the Josephon junction respectively, $\xi$ is the coherence length of the normal metal.[13] In our experiment, the length of the junction (or the wire) is much longer than the diameter of the wire. It is interesting that even in the 10 μm Nb strip, the vortices appear to be roughly linearly spaced along the superconducting strip.[15]

In summary we show evidence that the periodic oscillations in d$R$/d$B$ observed in Au nanowires contacted by supercodnucting W electrdoes are the consequence of the sequential generation and moving of vortices across the wires. According to our analysis, the voltage drop staircase pattern of the S-NW-S junction is determined by the viscosity coefficient of vortices $\eta$ and the resistance of the nanowire in the normal state $R_n$.

**Acknowledgements**

We are grateful to Moses Chan and Jainendra Jain for helpful discussions. We thank Bangzhi Liu for the help in TEM study. This work was financially supported by National Natural Science Foundation of China (No. 11004010), the Fundamental Research Funds for the Central Universities, and the Penn State MRSEC under NSF grant DMR-0820404.

**Competing financial interests**

The authors declare that they have no competing financial interests.

Figure Caption

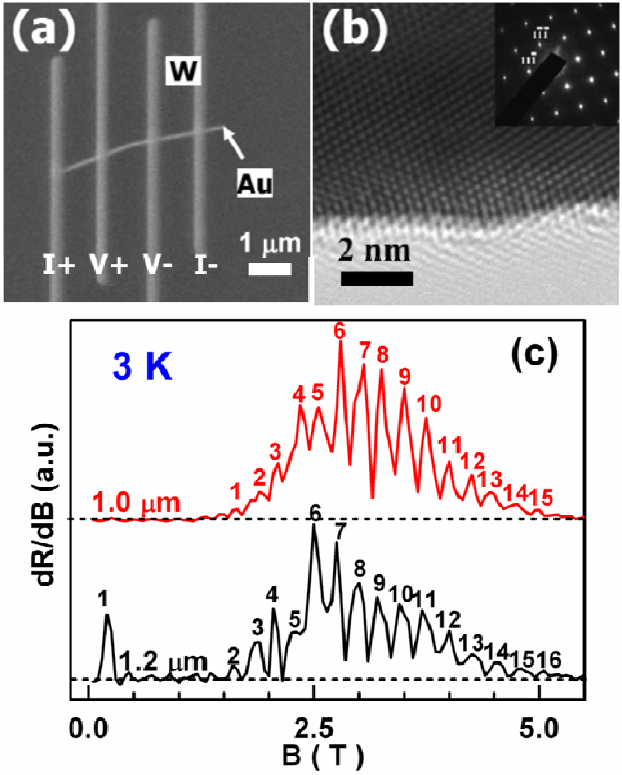

FIG. 1. (a) A scanning electron micrograph of the Au nanowire contacted by four FIB-deposited superconducting W electrodes. The magnetic field is applied perpendicular to the plane. (b) The high-resolution transmission electron microscope image and the corresponding SAED patterns of the Au nanowire. (c) The differential magnetoresistance as a function of magnetic field of the 1.0 μm and 1.2 μm Au nanowires measured at $T$ = 3 K (reproduced with magnification from reference 18). The dashed lines represent $dR/d|B|$ = 0 for the two wires. There are small oscillations of the value $dR/d|B|$ around zero possibly due to the limited resolution in $R$ and $B$ readings in low magnetic field. We picked the clearly resolved peak at about 1.65 T as the first peak of the 1.0 μm wire. The peak at about 1.6 T of the 1.2 μm Au nanowire was numbered as the $N$ = 2 peak. Except for the first peak of 1.2 μm nanowire, the differential magnetoresistance shows uniform oscillations with $B$ of 0.25 T.

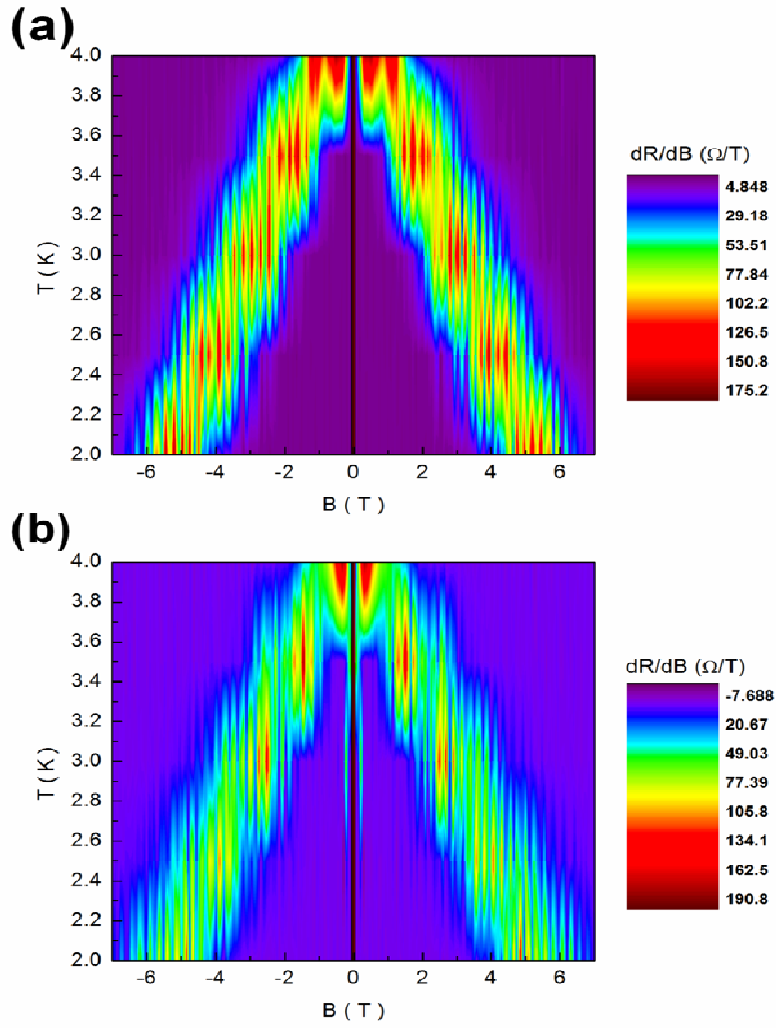

FIG. 2. The differential magnetoresistance as a function of magnetic field of the (a)1.0 μm and (b)1.2 μm Au nanowires measured at various temperatures. The color scales as *dR/dB*.

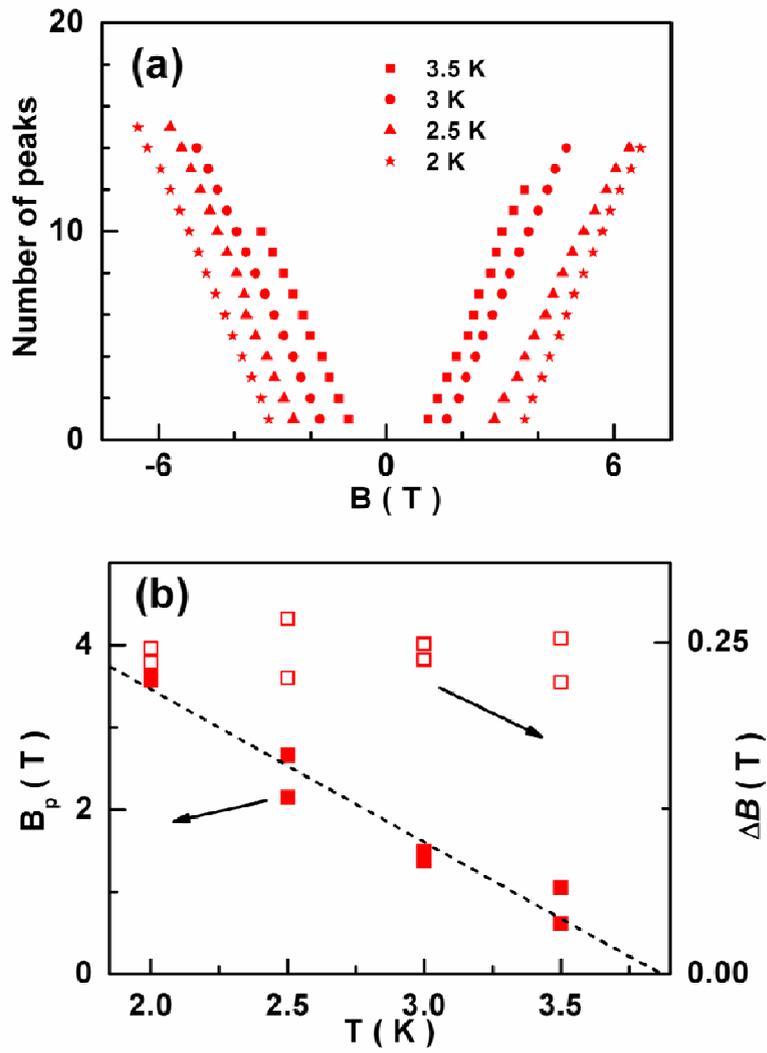

FIG. 3. (a) Peak number of $dR/d|B|$ curves as a function of applied magnetic field of the 1.0 μm nanowire measured at $T$ = 2 K, 2.5 K, 3 K, and 3.5 K. (b) The 'critical' field $B_p$ (solid symbols) and the slopes $\Delta B$ (open symbols) as a function of temperature obtained according to Eq. (1). The dashed line is guide to eyes.

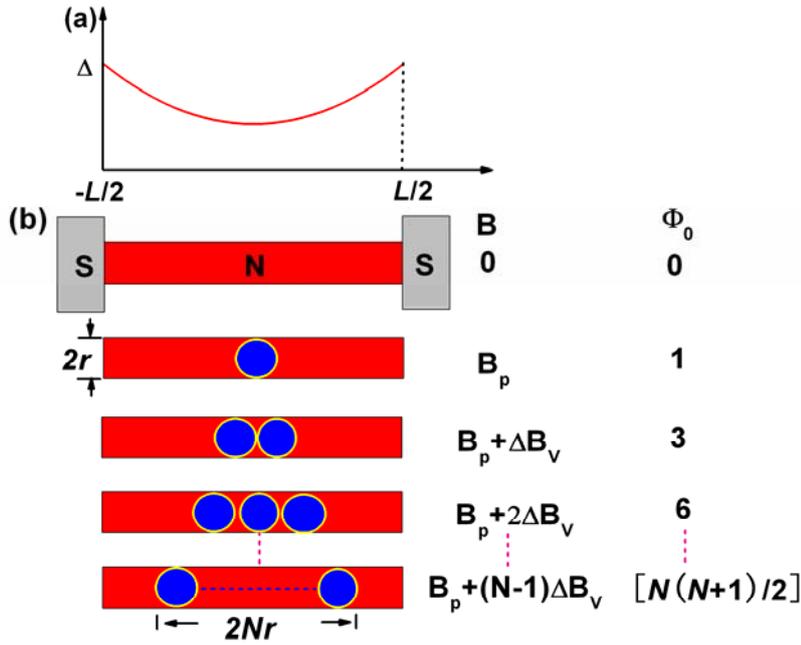

FIG. 4. (a) The superconducting gap as a function of the position induced by the proximity effect in the nanowire. (b) Schematic view of the S-NW-S structure and the vortices induced by the applied field. Below the critical field $B_p$, all flux was expelled from the nanowire. At $B_p$, one vortex carrying one flux quantum $\Phi_0$ enters the nanowire. Above this critical field, additional vortices are induced one at a time with the magnetic field widths of $\Delta B_V = \Phi_0/2\pi r^2$.

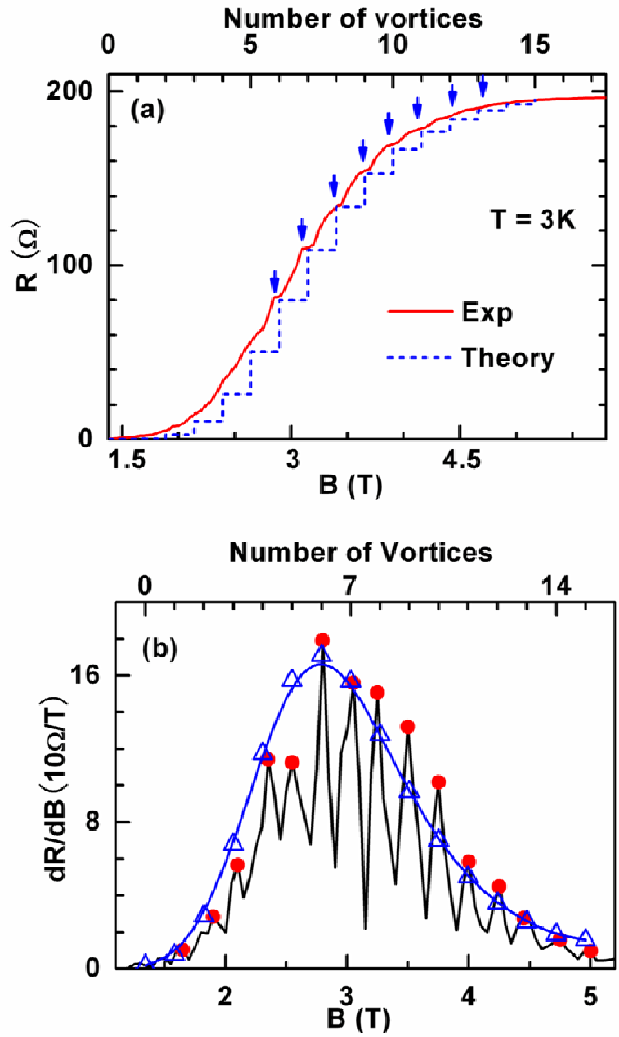

FIG. 5. (a) The calculated resistance $R_N = V_N/I$ by taking into account $a = 14$ $\Omega^{-1}$ and $R_n = 197$ $\Omega$ and the experimental result measured at 3 K. The arrows point to the apparent resistance staircases in high magnetic field. (b)The height of the peaks in the $dR/d|B|$ curves measured at 3 K (●) and the $dR/d|N|$ calculated by the theoretical curve in Fig. 5(a) (△) as a function of the number of vortices of the 1.0 μm wire. The solid curve is the guide to eyes.

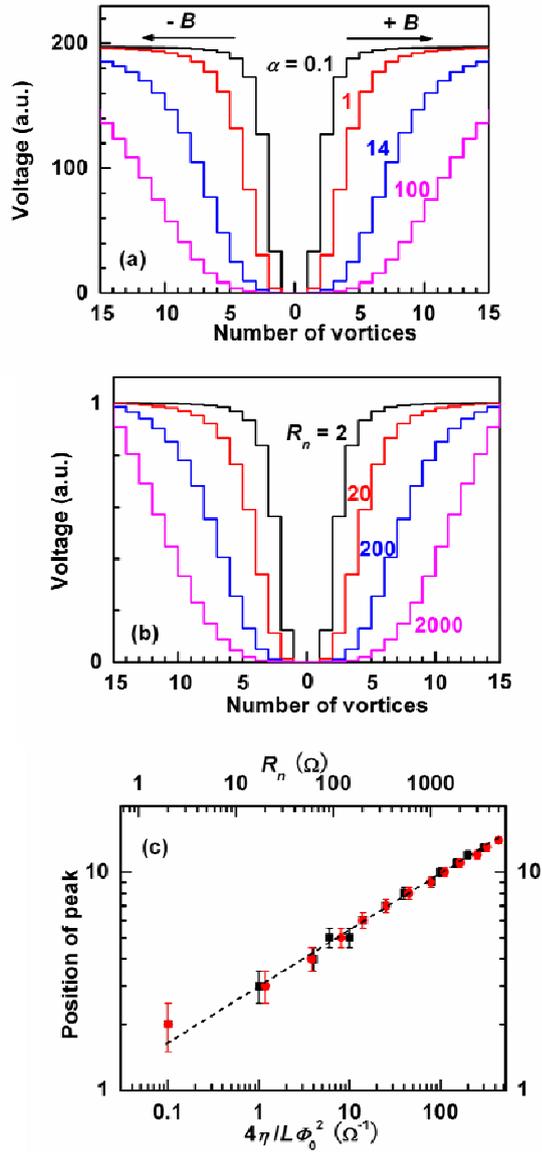

FIG. 6. The dependence of voltage drop on the number of vortices (or on the magnetic field) in the S-NW-S junction calculated by Eq. (4). (a) The calculated voltage drop as a function of the number of vortices calculated with $R_n$ = 197 Ω and $a = 4\eta/(L\Phi_0^2)$ = 0.1, 1, 14, and 100 Ω$^{-1}$. (b) The calculated voltage drop as a function of the number of vortices calculated with $a = 4\eta/(L\Phi_0^2)$ = 14 Ω$^{-1}$ and $R_n$ = 2, 20, 200, and 2000 Ω. (c) The dependence of the position of peak in $dV/dN$ curves on the parameters $R_n$ and $a = 4\eta/(L\Phi_0^2)$ of the S-NW-S junction. The solid cycles (●) are determined by fixing $a$ = 14 Ω$^{-1}$ and varying $R_n$ from 2 to 5000 Ω. The solid squares (■) are determined by fixing $R_n$ = 197 Ω and varying $a$ from 0.1 to 300 Ω$^{-1}$. The dashed line is the guide to eyes.

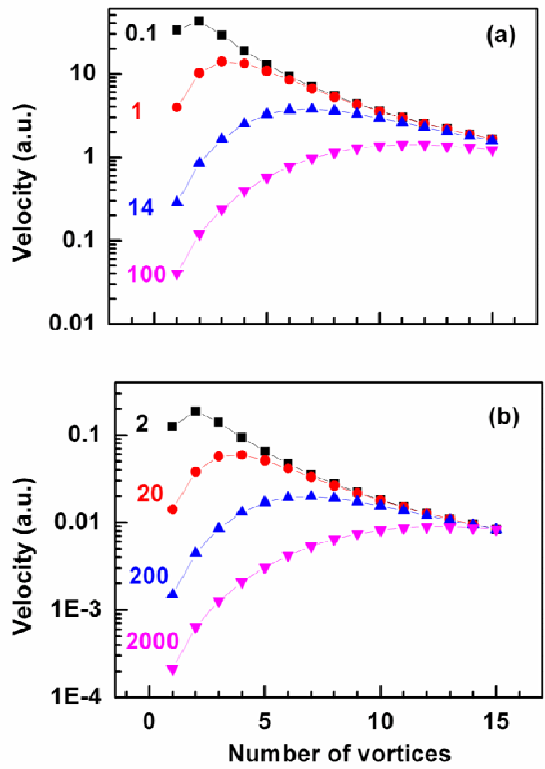

FIG. 7. The dependence of vortices velocity on the number of vortices in the S-NW-S junction. (a) The calculated vortices velocity as a function of the number of vortices calculated with $R_n$ = 197 Ω and $a = 4\eta/(L\Phi_0^2)$ = 0.1, 1, 14, and 100 Ω$^{-1}$. (b) The calculated vortices velocity as a function of the number of vortices calculated with $a = 4\eta/(L\Phi_0^2)$ = 14 Ω$^{-1}$ and $R_n$ = 2, 20, 200, and 2000 Ω.